\documentclass{IEEEoj}
\usepackage{cite}
\usepackage{amsmath,amssymb,amsfonts}
\usepackage{algorithmic}
\usepackage{graphicx,color}
\usepackage{textcomp}
\usepackage{placeins}
\usepackage[none]{hyphenat} 
\usepackage{amsmath, amssymb}
\usepackage{stfloats} 
\usepackage{cuted}    
\usepackage{bm} 
\def\BibTeX{{\rm B\kern-.05em{\sc i\kern-.025em b}\kern-.08em
    T\kern-.1667em\lower.7ex\hbox{E}\kern-.125emX}}
\def\OJlogo{\vspace{-10pt}\includegraphics[height=20pt]{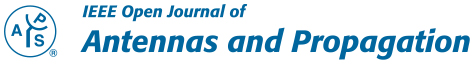}}
\begin{document}
\receiveddate{XX Month, XXXX}
\reviseddate{XX Month, XXXX}
\accepteddate{XX Month, XXXX}
\publisheddate{XX Month, XXXX}
\currentdate{XX Month, XXXX}
\doiinfo{OJAP.2024.1234567}

\title{Near-Field Collinear Dipole Array Design with Dual-Polarization Operation for Wireless Power Transfer}

\author{Jiawang Li\authorrefmark{1}}
\affil{Department of Electrical and Information Technology, Lund University, Lund, Sweden}
\corresp{CORRESPONDING AUTHOR: Jiawang Li (e-mail: jiawang.li@eit.lth.se).}
\markboth{Preparation of Papers for \textsc{IEEE Open Journal of Antennas and Propagation}}{Jiawang Li}

\begin{abstract}
A large intelligent surface (LIS) is a promising approach to enhancing 6G performance in sub-10 GHz frequency bands. An analysis of the horizontal dipole linear array reveals that when the array size is sufficiently large, employing the conjugate phase method to excite the antenna elements results in a focal point along the central line of the array that gradually stabilizes at a constant value. The study evaluates the 3dB focal resolution generated by the dipole array for two polarizations. Additionally, the feasibility of generating circularly polarized focal points and the focal displacement under a limited array size are explored.
Previous near-field focal synthesis methods primarily considered only the line-of-sight (LOS) channel. To enhance practicality, this study adopts a two-ray channel model to analyze near-field focal synthesis results in the presence of a reflected path. Both two polarizations are discussed separately, with design guidelines provided for array placement and focal point positioning.
Finally, electromagnetic simulations are conducted within the linear array to validate the proposed design. These simulations highlight the capability of the horizontal dipole array configuration to be directly applied without significant coupling effects between elements. The proposed design guidelines lay a solid foundation for the application of LIS in 6G technology.
\end{abstract}


\maketitle

\section{INTRODUCTION}
\IEEEPARstart{T}{he} large intelligent surface (LIS) [1] is an innovative solution poised to meet the demanding requirements of 6G in the sub-10 GHz frequency bands, addressing the constraints of limited spectrum resources. Unlike phased array systems, which are typically analyzed in the far-field [2]-[3], LIS is predominantly deployed in indoor environments and residential areas, comprising numerous antenna elements distributed widely across the space. Therefore, LIS can be regarded as an evolution of the localized massive MIMO concept introduced in 5G.  
Since the user equipment operates in the near-field region, the transmission differs from the plane waves typically observed in the far field [4]-[5]. The impact of spherical wave propagation in the near field must be carefully evaluated [6]-[8]. 

Whether LIS is employed in massive MIMO communication or power transmission [9]-[13], it is essential to focus the beam energy to boost the signal-to-noise ratio or increase transmission efficiency.
Beamforming is the primary technique for concentrating beam energy, and the implemented methods in the radiative near field can be categorized into two main types [14]-[23]. The first method is the conjugate-phase approach [14]-[19], which has a clear physical interpretation: different phases are assigned to each element to compensate for delays caused by the physical propagation path. For instance, a multifocus antenna array with specified amplitude and phase conditions in the near-field region is developed, utilizing a modeling algorithm that incorporates phase compensation, superposition strategies to enhance spatial resolution and meet diverse multifocus requirements [17]. Besides, a curved substrate integrated waveguide (SIW) is utilized as a feeding carrier to enhance the amplitude and phase control capability of the near-field-focused (NFF) beam, enabling stable focal length and a wide steerable range [19]. 
The second method involves optimizing both the amplitude and phase of the antenna elements through brute-force search, typically employed in scenarios with specific objectives, such as maximizing transmission efficiency or designing flat-top beams [20]-[22]. Such as a near-field sparse array synthesis method based on Bayesian compressive sensing (BCS) and convex optimization is proposed to generate reference-shaped beams with controllable sidelobe levels, demonstrating effectiveness with over 50\% element reduction while maintaining desired near-field patterns [22]. The drawbacks include a lack of physical insight, potential slow convergence, and signal distortion caused by scattering in real-world channels. Therefore, this paper adopts the conjugate- phase method.

Moreover, circular polarization (CP) is frequently discussed in far-field applications, such as satellite communications [23] and WiFi systems [24]. Although the concept of circular polarization also exists in the near field, its application is predominantly limited to manipulating circular aperture [25]-[26]. Therefore, this paper further explores the possibility of using a linear array to generate a CP near-field focal point, aiming to address scenarios where the user remains stationary while exhibiting rotational or micro-motion. Unlike studies [10]-[13], which only consider line-of-sight (LOS) scenarios, this paper further investigates the impact of reflected paths in non-line-of-sight (NLOS) environments on both polarization types. It also provides installation guidelines to mitigate the effects caused by reflective paths.

\begin{figure}
	\centerline{\includegraphics[width=3.5in]{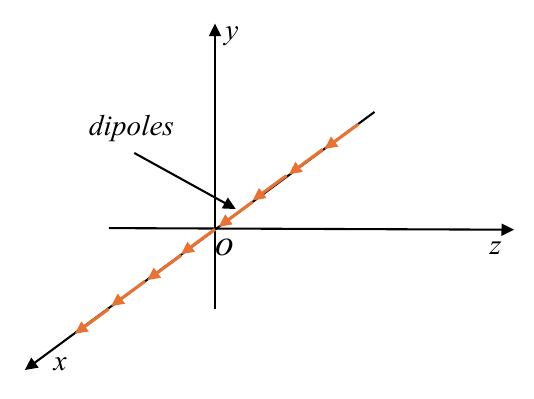}}
	\caption{Antenna array equipment for collinear dipoles array.\label{fig1}}
 \end{figure}

\section{ANTENNA MODEL}
The arrangement of dipole arrays is typically classified into three types: side-by-side, collinear, and parallel-in-echelon [27]. Among these, the parallel-in-echelon configuration can be considered a hybrid mode that lies between the first two. Consequently, the side-by-side and collinear arrangements are more commonly employed. According to reference [27], the mutual impedance between dipoles in these two configurations can be calculated using the following:
\begin{equation}
\begin{aligned}
Z_{ss} = & \frac{\eta}{4\pi} \big[ 2C_i(u_0) - C_i(u_1) - C_i(u_2) \\
& - j \big( 2S_i(u_0) - S_i(u_1) - S_i(u_2) \big) \big]
\end{aligned}
\end{equation}

\begin{equation}
\begin{aligned}
Z_{co} = & \frac{\eta}{8\pi} \big[ \big( 2S_i(2v_0) - S_i(v_2) - S_i(v_1) \big) 
\big( \sin(v_0) - j\cos(v_0) \big) \big] \\
& + \frac{\eta}{8\pi} \big[ 
j \sin(v_0) \big( 2C_i(2v_0) - C_i(v_2) - C_i(v_1) - \ln(v_3) \big) \\
& \quad - \cos(v_0) \big( C_i(v_2) - 2C_i(2v_0) + C_i(v_1) - \ln(v_3) \big) 
\big]
\end{aligned}
\end{equation}

Compared to side-by-side arrays, collinear arrays exhibit lower coupling between antenna elements, allowing the theoretical model to achieve closer agreement with electromagnetic simulation results. As shown in Fig.1,
The horizontal polarization is chosen for each element, that is means, the direction of the electric field (E-field) of the antenna is parallel with the $\it{x}$-axis. We apply several simplifying assumptions of the array elements: no mutual coupling and only consider the LOS channel at the beginning. All the simulation results are based on the 6 GHz. To demonstrate the impact of horizontal polarization more clearly, here we use a horizontal dipole as an example and borrow the far-field propagation formula of a dipole antenna. In the far-field region, the vector potential $\boldsymbol{A}$ of a Hertzian dipole is [27]:

\begin{equation}
\boldsymbol{A}(\boldsymbol{r}) = \frac{I_0 \mu_0 \boldsymbol{l}}{4 \pi r} e^{-jkr}
\end{equation}

\begin{figure}
	\centerline{\includegraphics[width=3.5in]{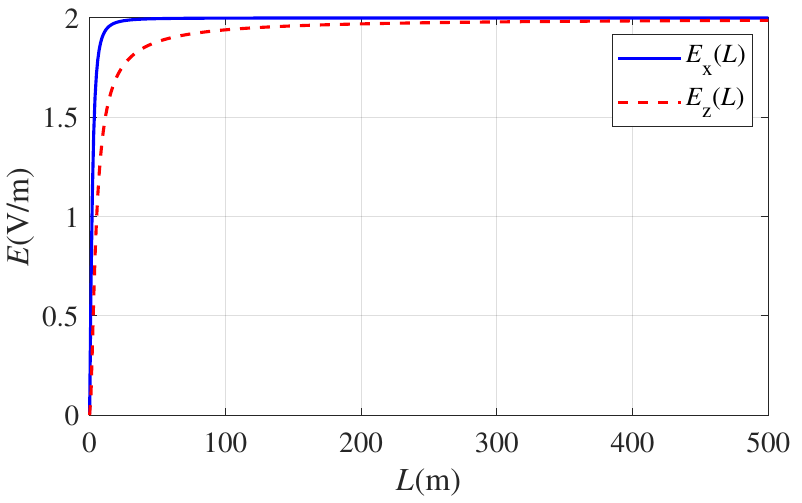}}
	\caption{The variations of $E_x$ and $E_z$ with respect to the the array length $L$ under a fixed focal position $z_o$ = 1.5m.\label{fig2}}
 \end{figure}
where \( r \) is the distance from the dipole to the field point, \( l \) is the infinitesimal dipole’s length,  
and \( \mu_0 \) is the permeability in vacuum. The E-field $\boldsymbol{E}$ is obtained from the time derivative of the vector potential: 
\begin{equation}
\bm{E} = -j\omega \bm{A} - \nabla \phi = -j\omega \bm{A} - j \frac{1}{\omega \mu_0 \varepsilon_0} \nabla (\nabla \cdot \bm{A}),
\end{equation}
Considering the polarization of the dipole, and assuming $I_0$ is a constant value, the final $\boldsymbol{E}$ can be expressed as:
\begin{equation}
	\boldsymbol{E}(\boldsymbol{r},\boldsymbol{l})=\frac{j\eta I_0\boldsymbol{l}k}{4\pi r}e^{-jkr}(\hat{r}\times(\hat{l}\times\hat{r})),
\end{equation}
 $\eta$ is the wave impedance in the free space. $\hat{\boldsymbol{l}}$ is the unit vector along the current of the dipole. The linear array is composed of $\it{N}$ antenna elements. $\emph{r}$ is the distance between the space field point and the center fo the dipole antenna. The total electric field at a certain point in space is:
 \begin{equation}
 	\boldsymbol{E}( \boldsymbol{r},\boldsymbol{l})=\sum_{i=0}^N\frac{j\eta I_0\boldsymbol{l}k}{4 \pi r_i}e^{-jkr_i}(\hat{r}_i\times(\hat{l}\times\hat{r}_i)),
 \end{equation}
Because the following relationship between  $\hat{{l}}$ and $\hat{r}$:
\begin{equation}
\hat{r}\times(\hat{l}\times\hat{r})=(\hat{r}\cdot \hat{r})\hat{l}-(\hat{r}\cdot\hat{l})\hat{r}=\hat{l}-(\hat{r}\cdot\hat{l})\hat{r},
 \end{equation}
 \begin{figure}
	\centerline{\includegraphics[width=3.5in]{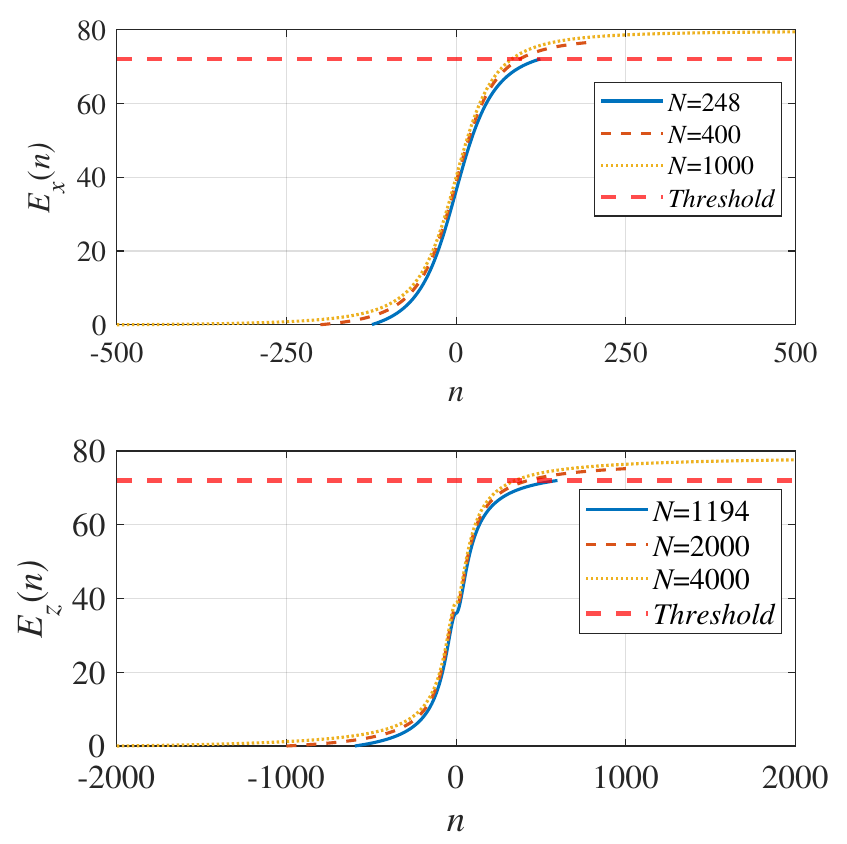}}
	\caption{Comparison Between the Numerical and Analytical Solutions of Peak E-Field Intensity. \label{fig3.pdf}}
 \end{figure}
 thus, the field point, source point, and dipole placement can be established in the Cartesian coordinate system to derive the total electric field.
The polarization of all the elements is along the $\it{x}$-axis, so  $\hat{l}$ can be expressed as (1,0). It is important to note that, although this article focuses on a near-field problem, the studied area is in the far-field region for each antenna element relative to the array. The results can therefore be obtained as follows:

\begin{figure}
	\centerline{\includegraphics[width=3in]{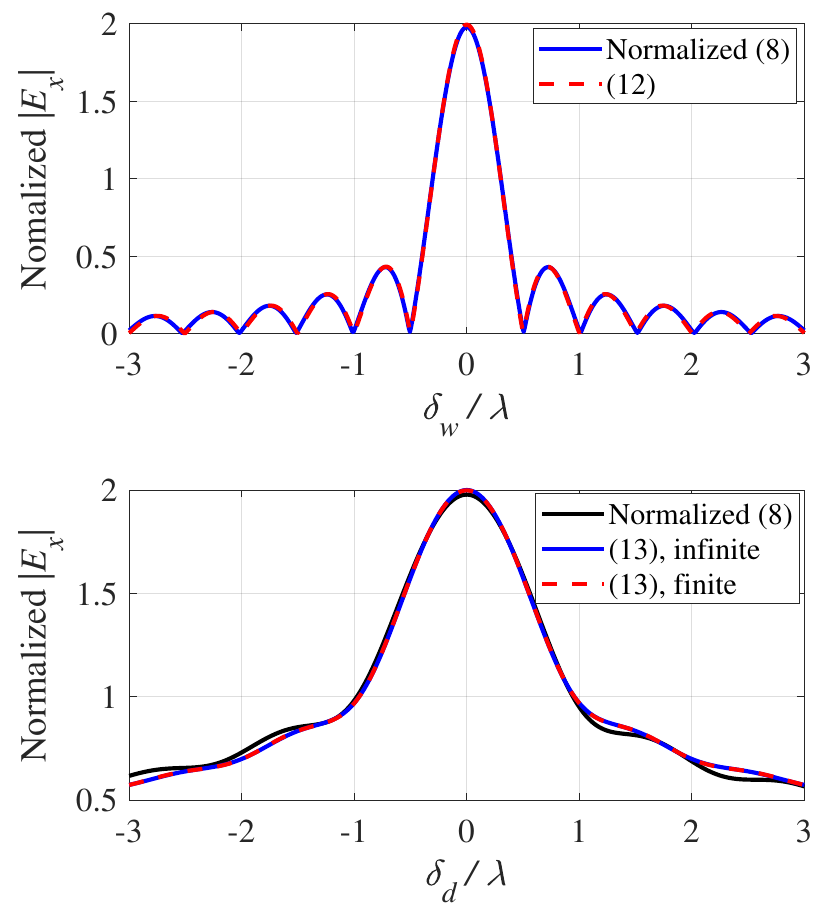}}
	\caption{Beamwidth and beamdepth for the $E_x$.\label{fig4}}
 \end{figure}
\begin{equation}
E_x(x, z) = \sum_{n=-N/2}^{N/2} \frac{ z^2 e^{-j k \sqrt{z^2 + (nd - x)^2}} }{ \left( z^2 + (nd - x)^2 \right)^{3/2} },
\end{equation}

\begin{equation}
E_z(x, z) = \sum_{n=-N/2}^{N/2} 
\frac{ (nd - x) z e^{-j k \sqrt{z^2 + (nd - x)^2}} }{ \left(z^2 + (nd - x)^2 \right)^{3/2} },
\end{equation}

It is important to note that when designing using the conjugate-phase method for (8) and (9), not only the phase differences caused by path variations among different elements should be considered, but also those induced by the symmetry of the array structure. Specifically, with respect to (9), the phase difference between symmetric elements, centered at the origin, along the positive and negative directions of the $x$-axis is $\pi$. 

When conjugate-phase method is applied solely to $E_x$, this results in $E_z = 0$. Conversely, after compensating for the phase difference introduced by the array's structural symmetry in $E_z$, $E_x$ becomes 0. This implies that adopting distinct conjugate-phase strategies for $E_x$ and $E_z$ can effectively achieve low cross-polarization levels at focal positions.
To facilitate the observation of the focusing E-field's variation with the focal point and array aperture, the element spacing can be assumed to be very small. This assumption allows for the approximation of a continuous aperture array with a length of \(L\) focused at the focal point \(z_0\). The following integral can then be applied to calculate this scenario:
 
\begin{equation}
E_x(z_o) = \int_{-L/2}^{L/2} \frac{z_o^2}{\left(z_o^2 + s^2\right)^{3/2}} ds
= \frac{2}{\sqrt{4\left(\frac{z_o}{L}\right)^2 + 1}},
\end{equation}
 
\begin{equation}
E_z(z_o) = \int_{-L/2}^{L/2} \frac{| s | z_o}{\left(z_o^2 + s^2\right)^{3/2}} ds 
= 2 \left( 1 - \frac{2}{\sqrt{4 + \left(\frac{L}{z_o}\right)^2}} \right).
\end{equation}
where $s$ is the distance offset the center of the linear array. As shown in Fig. 2, the plot illustrates the variation trend of the E-field intensity with array size $L$ when $z_o = 1.5$m, respectively. As the ratio $L/z_o$ increases, the values converge to a constant of 2. However, the convergence rate of $E_x$ is significantly faster than that of $E_z$. This implies that when the focal distance \( z_0 \) is a finite value, the E-field can remain constant within a certain distance. From the perspective of power, although it seems unreasonable that the power increases with the increase of distance, in fact, this is due to the continuous injection of energy from the sources existing in the array.

It can also be observed that when $Z_o = 1.5$m (a reasonable choice for certain narrow corridors, such as a 3-meter-wide corridor where dipole arrays can be placed on both sides), $E_x$ exceeds 1.9 when $L \geq 10$, whereas for $E_z$ to reach the same threshold, at least $L \geq 60$ is required. This implies that the ratio $L/z_o$ must be at least 6.67 and 40, respectively.

Furthermore, we also determined the precise number of antennas required at 6 GHz, given a half-wavelength spacing of $d = 0.025$m and $z_0 = 1.5$m, with the results presented in Fig. 3. As predicted, when the number of antennas becomes sufficiently large, both $E_x$ and $E_z$ asymptotically approach 80V/m. Here, we define the threshold as 90\% of the maximum E-field intensity, corresponding to over 81\% of the peak power. 

This threshold is a reasonable choice, as setting it above 90\% of the power would require the E-field intensity to reach at least 94.9\%. From the trend of the curves, it is evident that contributions from distant elements are minimal, leading to unnecessary energy consumption and increased costs. For $E_x$, its faster convergence allows it to reach the threshold with 248 antennas, corresponding to a physical size of 6.2 m. In contrast, for $E_z$, 1194 antenna elements are required to meet the threshold, resulting in a physical size of 29.85 m. When the array length is sufficiently large, the closed-form expressions for the beamwidth and beam depth can be derived as follows (these can be obtained from (32) and (33) in the Appendix):

\begin{equation}
E_x(\delta_w) \approx 2 \left| \sin \left( \frac{k\delta_w}{\sqrt{1 + 4\left(\frac{z_o}{L}\right)^2}} \right)/ (k\delta_w) \right| 
\approx 2 |\text{sinc}(k\delta_w)|
\end{equation}

\begin{equation}
E_x(\delta_d) \approx \pi \left| j J_1(k \delta_d) + H_{-1}(k \delta_d) \right| \quad L \to \infty
\end{equation}
The derived expressions for the beamwidth and beam depth of $E_x$ at the focal point are given in (12) and (13). Here, $\text{sinc}(\cdot)$ represents the sinc function, $J_1(\cdot)$ denotes the first-kind Bessel function, and $H_{-1}(\cdot)$ corresponds to the -1-order Struve function. The numerical solution obtained using (8) and the analytical solution derived from the formulas are shown in Fig. 4, demonstrating a high degree of agreement.

Furthermore, since (13) is derived under the assumption that $L$ approaches infinity, truncating it by setting the upper integration limit to a finite value close to $L = 40$ still provides a good approximation. This is because both $J_1(\cdot)$ and $H_{-1}(\cdot)$ exhibit oscillatory decay as the independent variable increases. The 3 dB beamwidth is found to be $0.21\lambda$, while the 3 dB beam depth is $0.6\lambda$. The beam depth is slightly greater than half a wavelength, indicating that the energy can be focused to cover most common antenna types, such as half-wave dipoles and patch antennas. Similarly, the closed-form expressions for the beamwidth and beam depth of \(E_z\) are as follows (these can be obtained from (34) and (35) in the Appendix):

\begin{equation}
E_z(\delta_w) \approx 
\pi\left| H_{-1}(k\delta_w)
\right|,
\quad L \to \infty
\end{equation}

\begin{equation}
E_z(\delta_d) \approx 
\frac{2 (z_0 + |\delta_d|)}{z_0} 
\left| \frac{i \left(1 - e^{i \delta k}\right)}{\delta k} \right|, 
\quad L \to \infty
\end{equation}

\begin{figure}
	\centerline{\includegraphics[width=3in]{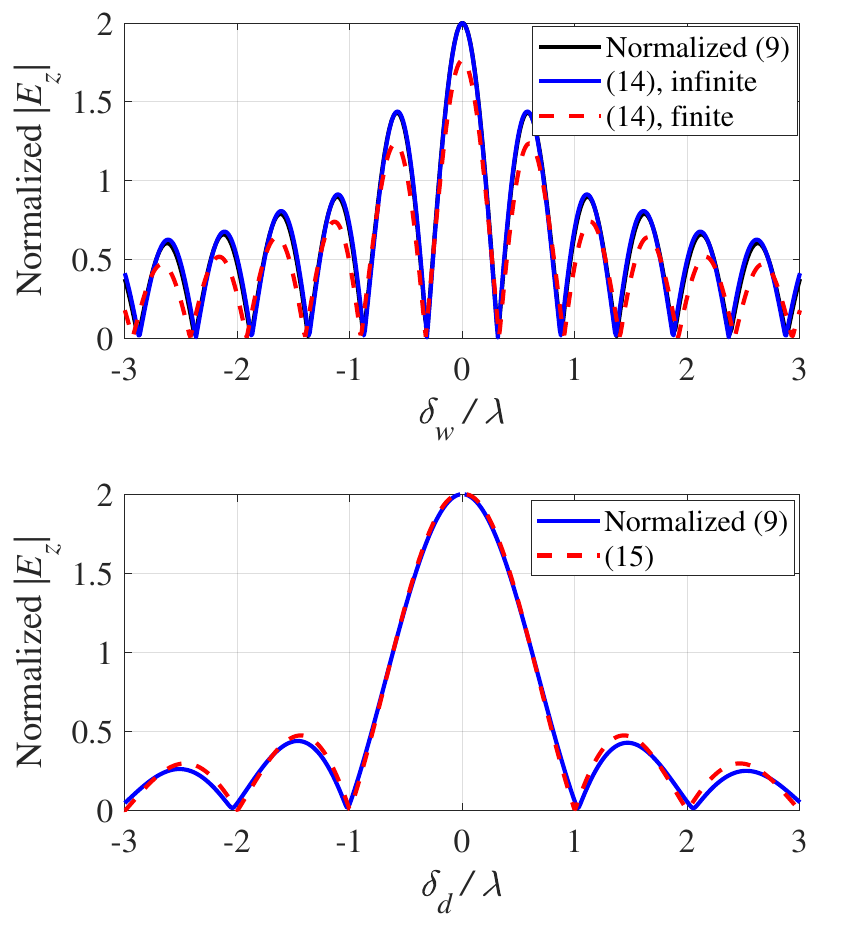}}
	\caption{Beamwidth and beamdepth for the $E_z$. \label{fig6}}
\end{figure}
As shown in Fig. 5, the variation in \(E_z\)-polarization E-field width and depth is presented. Unlike \(E_x\)-polarization, the 3 dB beamwidth and 3 dB beam depth of \(E_z\)-polarization are both narrower, measuring \(0.31\lambda\) and \(0.89\lambda\), respectively, with a ratio of approximately 2.87. Additionally, the sidelobe level in the beamwidth direction is relatively higher, with the strongest sidelobe being only about 3 dB lower than the main lobe. Methods such as Taylor beam-pattern synthesis can be employed to reduce the sidelobe level [28].

\section{CP BEAM GENERATION}
 A large number of CP antennas operating in the far-field have been reported in [23]-[24]. However, there are few reports on circularly polarized beams in the near-field [25]-[26]. This is mainly because the amplitudes of the two orthogonal polarizations are easily affected by the near-field distance, resulting in a significant difference in amplitude between the two polarizations. 
 In this design, since the electric field phases of \( E_x \) and \( E_z \) are already orthogonal, the subsequent analysis will mainly focus on the amplitudes of the E-fields. Two linear arrays of equal length are selected for the analysis. The distance between the two linear arrays is so small that they are nearly coincident, which is permissible due to the extremely small radial dimension of the dipoles. 
 \begin{figure}
	\centerline{\includegraphics[width=3in]{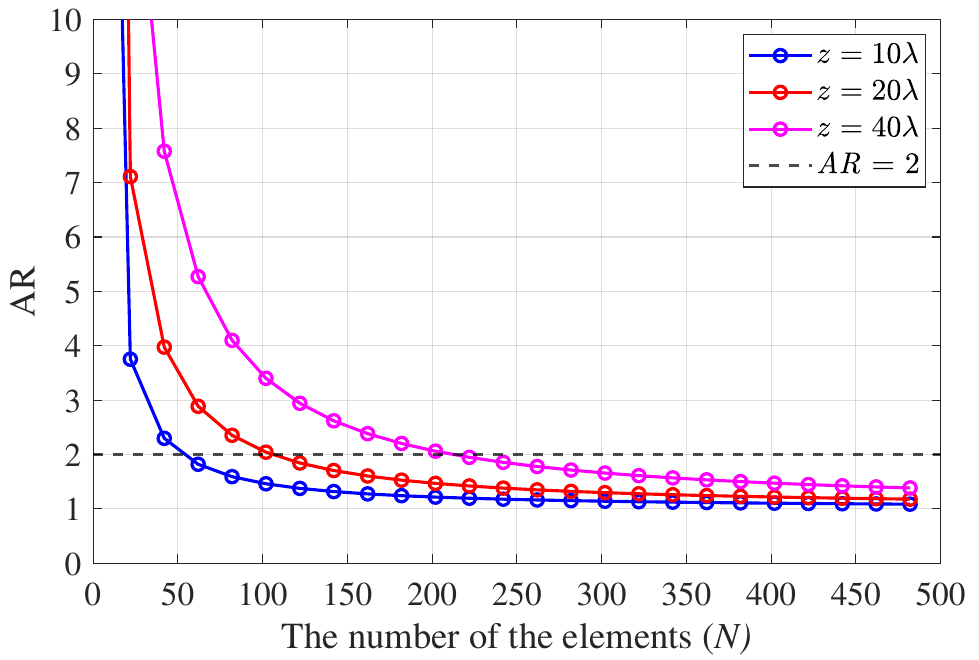}}
	\caption{AR vs the number of the elements. \label{fig7}}
\end{figure}
Moreover, the cross-polarizations of the two orthogonal polarizations at the central position are both 0. Therefore, we only need to consider the amplitudes of the two main polarizations, and assume that the distances from any point in space to the central positions of the two linear arrays are equal. Referring to the requirements for CP antennas in the far field [29], the antenna's axial ratio must meet specific criteria as following:
\begin{equation}
20 \log_{10} (AR) \leq 3\text{dB} \Rightarrow AR \leq 2
\end{equation}
This implies that ($E_x \geq E_z$ with the same elements):
\begin{equation}
1 \leq \frac{E_x}{E_z} \leq 2
\end{equation}
\begin{figure*}[htbp] 
    \centering
    \includegraphics[width=\textwidth]{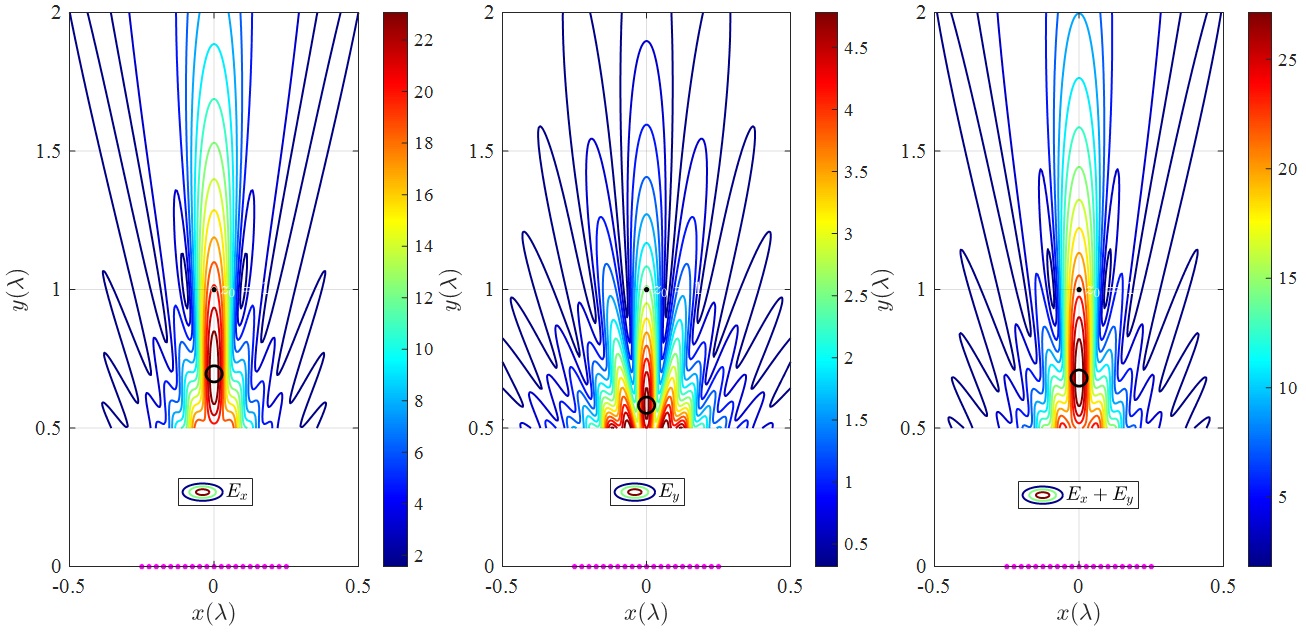}
    \caption{Electric field distribution using 20 elements.}
    \label{fig:electric_field}
    \centering
    \includegraphics[width=\textwidth]{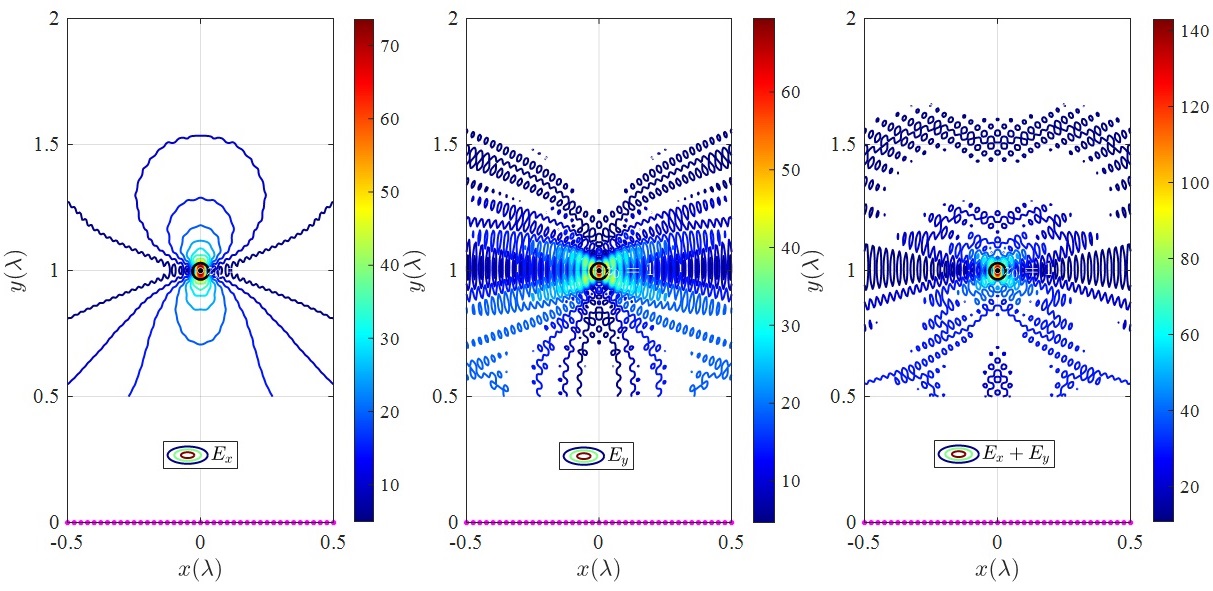}
    \caption{Electric field distribution using 2000 elements.}
    \label{fig:electric_field}
\end{figure*}
As shown in Fig. 6, the curves of the axial ratio versus the number of antenna elements are presented here for the cases where the focal positions are at \(10\lambda\), \(20\lambda\), and \(40\lambda\) (0.5m, 1m, 2m, respectively). CP focal regions can be generated when the number of elements reaches at least 53, 106, and 213 respectively. An interesting observation is that when the focal distance is relatively short, the minimum number of antenna elements required to achieve CP is approximately proportional to the focal distance. This indicates that narrower corridors or rooms require fewer antenna elements. Considering that the actual width of a typical long corridor is limited, for wider corridors, antennas can be placed on the two side walls respectively, which can halve the focusing distance. Another reason these two polarizations can synthesize circular polarization is the slight difference in their 3dB beam resolution.

As shown in Fig. 7 and Fig. 8, the distributions of the E-field when the number of antenna elements for two polarizations is 20 and 2000 respectively, with the designed focal point located at (0, 20$\lambda$), and the schematic diagram of the total E-field distribution generated by the superposition of the two are presented here. The black dot represents the target focal position, the black circle represents the actual position of the focusing peak, and the pink dots on $z=0$ represent the positions of the antenna elements. When the number of antennas is relatively small, equal to 20, a significant focal shift phenomenon can be observed [16]. Moreover, the focal shift caused by $E_x$ is significantly smaller than that caused by $E_z$. 

This indicates that in addition to being affected by the size of the array, the focal shift is also influenced by the antenna polarization. When the number of antennas increases to 2000, the focal points generated by $E_x$ and $E_z$ at (0, 20$\lambda$) tend to be the same, which naturally leads to the shape of the focal point after their superposition being similar to the original one. 

Additionally, it can be seen that the E-field around $E_x$ is relatively 'cleaner', while there are more sidelobes near the focal point generated by $E_z$, especially along the direction of the focal beamwidth. Then, near the CP focal point, the sidelobes can be significantly improved, and the focal position becomes more concentrated.

\section{MULTIPATH EFFECT}

Most of the current work on near-field analysis only considers the LOS channel \([ \text{10} ]\)-\([ \text{13} ]\). In fact, due to the relatively short operating distance of the near-field, it is more applicable in short-distance environments such as indoors. 
However, affected by the reflection, diffraction, and scattering of the environment like walls, floors, and roofs, the actual focal position and shape in practice are not as perfect as in the ideal case, and may even lead to poor focusing effects. 
Therefore, it is actually necessary to consider the impact of channel characteristics on the focusing effect. The traditional two-ray channel model \([ \text{30} ]\) is applicable to the case where there is ground reflection. 
\begin{figure}
	\centerline{\includegraphics[width=2.6in]{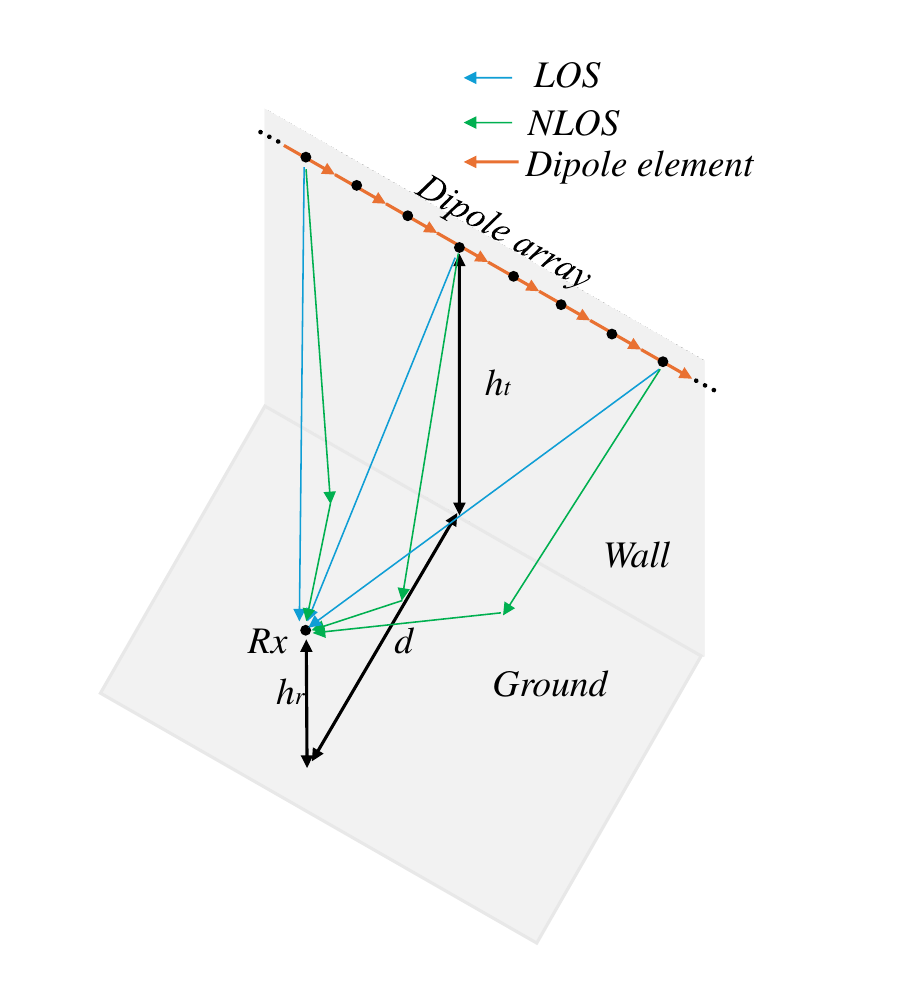}}
	\caption{Two-ray ground-reflection model in the near-field. \label{fig1}}
 \end{figure}
Thus, in our near-field analysis, we also consider using this model for analysis, as shown in Fig. 9.
From the Fig. 9, the received line of sight component can be written as:
\begin{equation}
    r_{los}^h(t) = \Re \left\{ \frac{\sqrt{G_{los}}}{2k} \times \frac{s(t) L_1^2 e^{-j k \sqrt{L_1^2+(nd)^2}}}{{\sqrt{L_1^2+(nd)^2}}^3} \right\}
\end{equation}
\begin{equation}
    r_{los}^v(t) = \Re \left\{ \frac{\sqrt{G_{los}}}{2k} \times \frac{s(t) nd L_1 e^{-j k \sqrt{L_1^2+(nd)^2}}}{{\sqrt{L_1^2+(nd)^2}}^3} \right\}
\end{equation}
and the ground reflected component can be written as:
\begin{equation}
\begin{split}
r_{gr}^{h}(t) &= \Re \left\{ 
\frac{\Gamma(\theta) \sqrt{G_{gr}}}{2k} \times 
\frac{s(t - \tau) (L_2' + L_2'')^2}
{\sqrt{(L_2' + L_2'')^2 + (nd)^2}^3} \right. \\
&\quad \times \left. e^{-jk \sqrt{(L_2' + L_2'')^2 + (nd)^2}}\right\}
\end{split}
\end{equation}
\begin{equation}
\begin{split}
r_{gr}^{v}(t) &= \Re \left\{ 
\frac{\Gamma(\theta) \sqrt{G_{gr}}}{2k} \times 
\frac{s(t - \tau) nd(L_2' + L_2'')}
{\sqrt{(L_2' + L_2'')^2 + (nd)^2}^3} \right. \\
&\quad \times \left. e^{-jk \sqrt{(L_2' + L_2'')^2 + (nd)^2}}\right\}
\end{split}
\end{equation}

where \( s(t) \) is the transmitted signal, $L_1$ is the length of the direct line-of-sight (LOS) ray for the center element, $L_2' + L_2''$ is the length of the ground-reflected ray for the center element, \( G_{los} \) is the combined antenna gain along the LOS path, and \( G_{gr} \) is the combined antenna gain along the ground-reflected path. $\Gamma(\theta)$ is the ground reflection coefficient and $\tau$ is the delay spread of the model, which equals:
\begin{equation}
(L_2' + L_2'' - L_1) / c.
\end{equation}
where $c$ represents the speed of light in a vacuum.
The ground reflection coefficient is:
\begin{equation}
    \Gamma(\theta) = \frac{\sin\theta - X}{\sin\theta + X}
\end{equation}
where $X = X_h$ or $X = X_v$, depending on whether the signal is horizontally or vertically polarized, respectively. $X$ is computed as follows:
\begin{equation}
    X_h = \sqrt{\varepsilon_g - \cos^2\theta}, \quad
    X_v = \frac{\sqrt{\varepsilon_g - \cos^2\theta}}{\varepsilon_g}
\end{equation}
The constant $\varepsilon_g$ is the relative permittivity of the ground. $\theta$ is the angle between the ground and the reflected ray, as shown in the figure above.

From the geometry of the figure, we get:
\begin{equation}
    L_2' + L_2'' = \sqrt{(h_t + h_r)^2 + L_g^2}
\end{equation}
\begin{equation}
    L_1 = \sqrt{(h_t - h_r)^2 + L_g^2},
\end{equation}
Therefore, the path-length difference between them is
\begin{equation}
    \Delta d = L_2' + L_2''- L_1 = \sqrt{(h_t + h_r)^2 + L_g^2} - \sqrt{(h_t - h_r)^2 + L_g^2}
\end{equation}
and the phase difference between the waves is
\begin{equation}
    \Delta \phi = \frac{2\pi \Delta d}{\lambda}
\end{equation}
From geometric relationships, the angle \( \theta \) can be determined as:
\begin{equation}
   \theta_n = \arctan \left( \frac{\sqrt{(h_t + h_r)^2+(nd)^2}}{\sqrt{L_1^2+(nd)^2}} \right)
\end{equation}
Therefore, the E-field expression at the receiving position corresponding to horizontal polarization, i.e., \( x \)-polarization. Since a dipole antenna exhibits omnidirectional radiation in the $\phi$-plane, it follows that $G_{\text{los}} = G_{\text{gr}}$. The amplitude is then normalized accordingly.

Figure~10 illustrates the distribution of the horizontal polarized E-field in a multipath environment. For simplicity, the antenna and focal positions are set at the same height. A significant difference in focal variation is observed when the focal height is set at $4\lambda$ compared to $40\lambda$. When the target focal point is located at $4\lambda$, two focal points are generated along the propagation direction. Due to focal displacement, the actual target focal point is positioned between $z = 10\lambda$ and $15\lambda$, while the focal point near $z = 20\lambda$ results from NLOS propagation. 

As the focal height increases, more focal points emerge along the propagation direction, indicating that under a limited array aperture, multipath effects lead to the degradation of near-field focal depth resolution. This suggests that if multiple users are arranged along the propagation direction, increased mutual interference may occur.

As shown in Fig. 11, when the number of elements increases to 2000, significant interference from the reflected path persists at shorter distances from the reflective surface (e.g., $h_t = 4\lambda$). This is primarily because the reflected path is relatively short, leading to less energy loss compared to the LOS path. However, increasing the height of the dipole array placement effectively suppresses multipath interference. Considering the dual reflection effects from both the floor and ceiling (assuming both surfaces have the same material properties), the optimal placement should be at the midpoint between the floor and ceiling to minimize NLOS energy propagation.

Additionally, Figure~12 presents the focal variation in the presence of a metallic floor. At shorter distances, the strong reflection properties of the metal cause the focal intensity of the reflected path to approach that of the LOS path. Although a clear focal point is still observed with increased array height, significant energy scattering occurs in the surrounding area.

Figure~13 presents the focusing results in a vertical polarized multipath environment. Unlike horizontal polarization, vertical polarization is less affected by multipath interference, resulting in a more stable focal region. This phenomenon is further demonstrated in Fig. 14, which shows that variations in height have a 
\begin{strip}
\begin{equation}
E_H =  \sum_{n=-N/2}^{N/2} \left[ 
\frac{\left((h_t - h_r)^2 + L_g^2\right) 
e^{-jk \sqrt{(h_t-h_r)^2 + L_g^2 + (nd-x)^2}}}
{\left((h_t-h_r)^2 + L_g^2 + (nd-x)^2\right)^{3/2}} 
+ \Gamma(\theta) 
\frac{\left((h_t + h_r)^2 + L_g^2\right) 
e^{-jk \sqrt{(h_t+h_r)^2 + L_g^2 + (nd-x)^2}}}
{\left((h_t+h_r)^2 + L_g^2 + (nd-x)^2\right)^{3/2}} 
\right] 
\end{equation}
\begin{equation}
E_V = \sum_{n=-N/2}^{N/2} \left[ 
\left|n\right| d 
\frac{\sqrt{(h_t-h_r)^2 + L_g^2} 
e^{-jk \sqrt{(h_t-h_r)^2 + L_g^2 + (nd-x)^2}}}
{\left((h_t-h_r)^2 + L_g^2 + (nd-x)^2\right)^{3/2}}+ \Gamma(\theta) \frac{\left|n\right| d 
\sqrt{(h_t+h_r)^2 + L_g^2} 
e^{-jk \sqrt{(h_t+h_r)^2 + L_g^2 + (nd-x)^2}}}
{\left((h_t+h_r)^2 + L_g^2 + (nd-x)^2\right)^{3/2}}
\right] 
\end{equation}
\end{strip}
minimal impact on vertical polarization, allowing it to maintain distinct focal characteristics consistently. 
Even under the conditions of a metallic reflective surface, as illustrated in Fig. 15, the energy dispersion in the focal region is significantly less pronounced than in the case of horizontal polarization E-fields.
\begin{figure}
	\centerline{\includegraphics[width=3.6in]{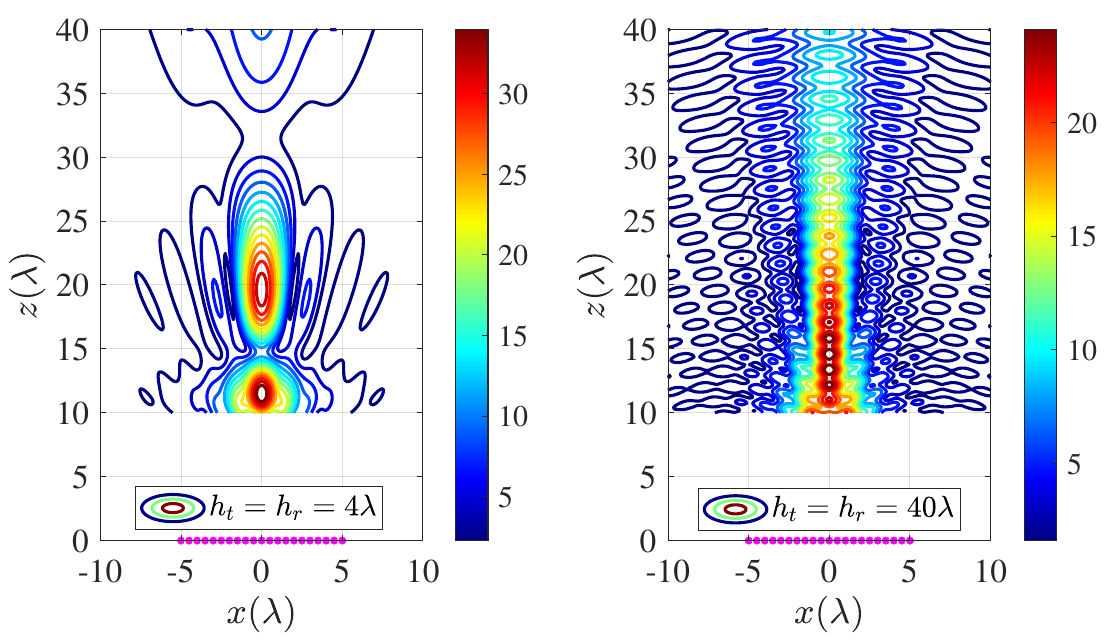}}
	\caption{$\it{E_h}$ under the element number of 20, $\varepsilon_g$ = 5 and focal point is at $20\lambda$. \label{fig7}}
	\centerline{\includegraphics[width=3.6in]{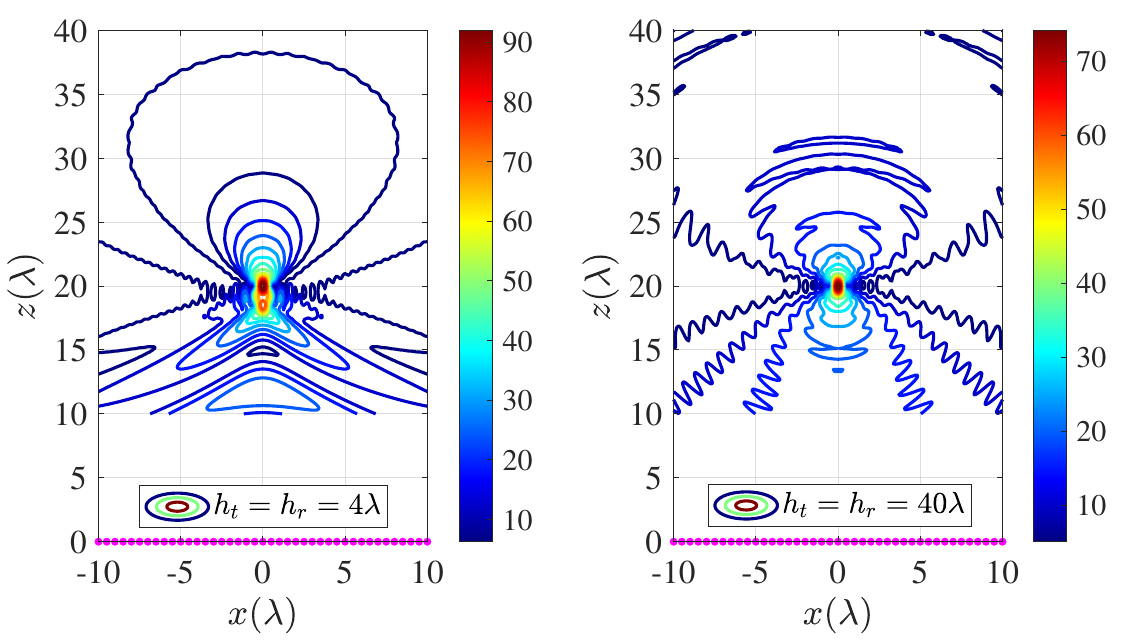}}
	\caption{$\it{E_h}$ under the element number of 2000, $\varepsilon_g$ = 5, and focal point is at $20\lambda$. \label{fig7}}
	\centerline{\includegraphics[width=3.6in]{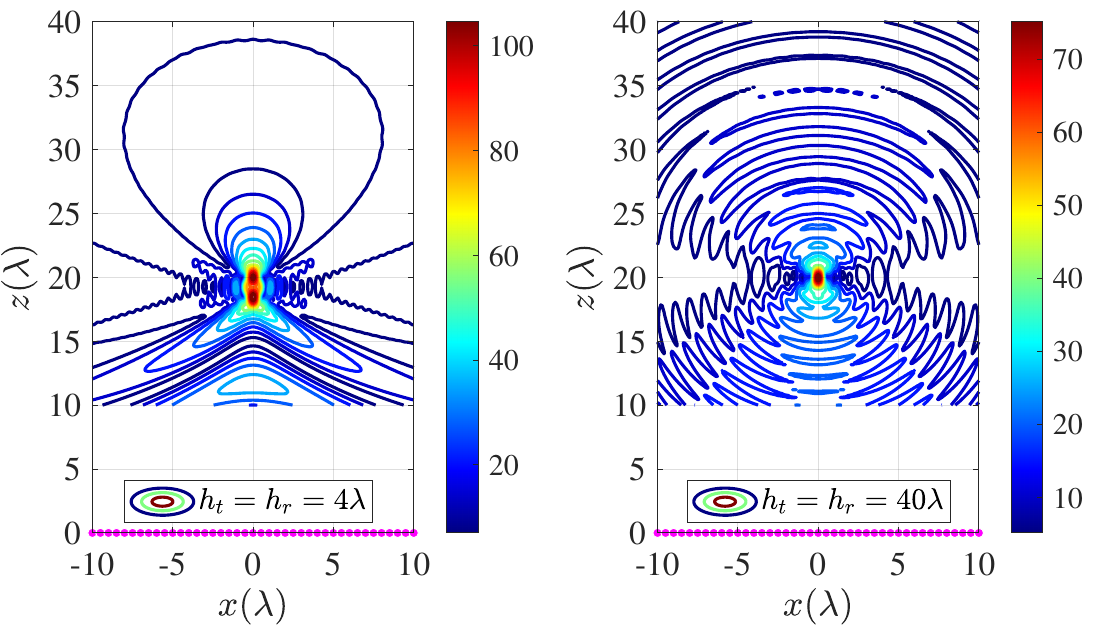}}
	\caption{$\it{E_h}$ under the element number of 2000, metal and focal point is at $20\lambda$. \label{fig7}}
\end{figure}
\begin{figure}
	\centerline{\includegraphics[width=3.6in]{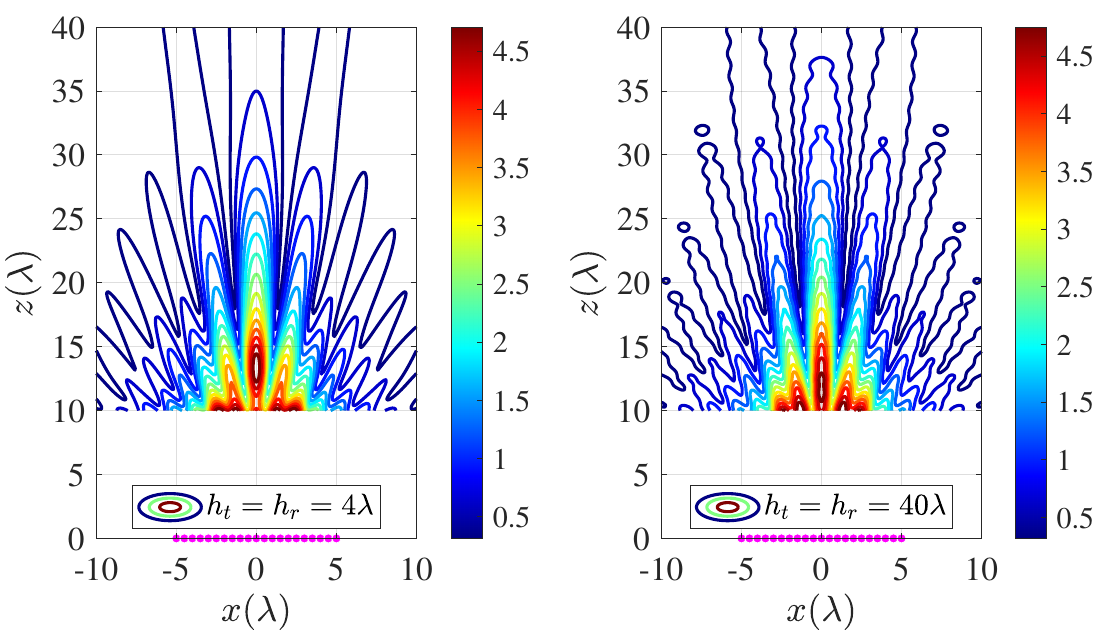}}
	\caption{$\it{E_v}$ under the element number of 20, $\varepsilon_g$ = 5 and focal point is at $20\lambda$. \label{fig7}}
    \centerline{\includegraphics[width=3.6in]{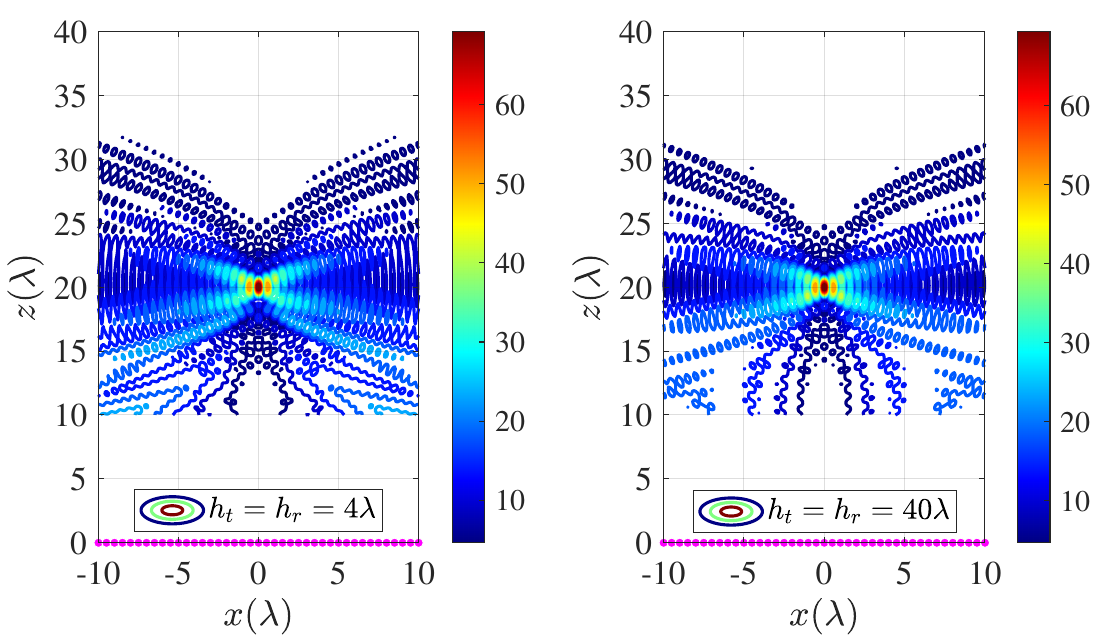}}
	\caption{$\it{E_v}$ under the element number of 2000, $\varepsilon_g$ = 5 and focal point is at $20\lambda$. \label{fig7}}
	\centerline{\includegraphics[width=3.6in]{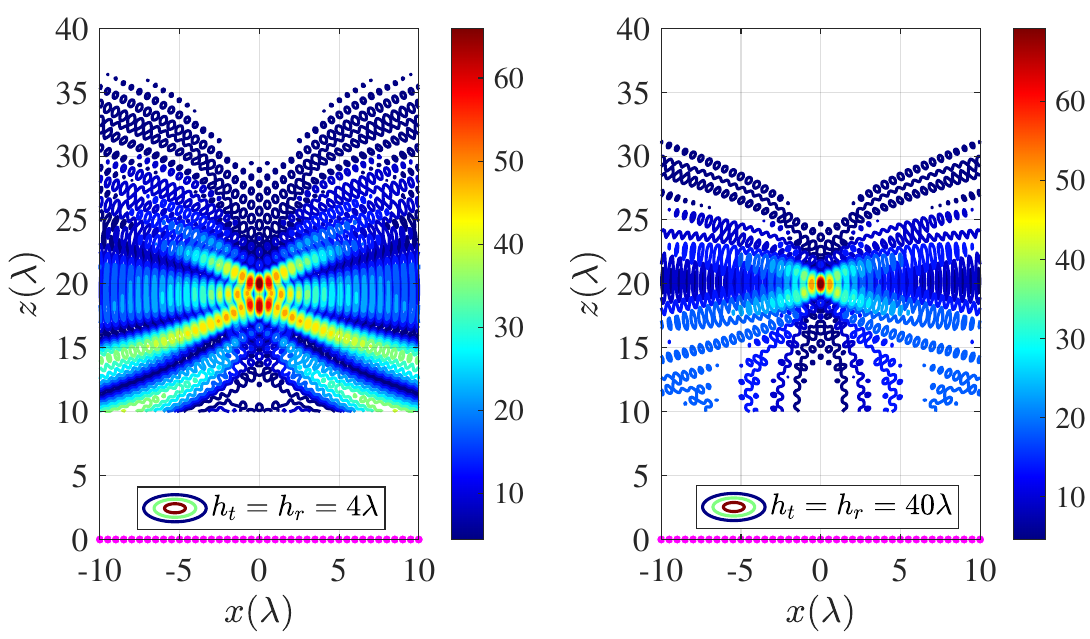}}
	\caption{$\it{E_v}$ under the element number of 2000, metal and focal point is at $20\lambda$. \label{fig7}}
\end{figure}

\section{ELECTROMAGNETIC SIMULATION VERIFICATION}
To evaluate the focusing performance of the dual-polarized focusing array proposed in this paper, the FEKO software is used to validate the array's performance in the $\it{Ex}$ and $\it{Ez}$ components. The specific process is as follows: MATLAB was employed to generate the FEKO script for the Hertzian dipoles, allowing precise control of the amplitude and phase of the dipoles. To closely approximate the ideal Hertzian dipole, a cylindrical model with a length of 0.1 mm and a radius of 0.01 mm is used to ensure a uniform current distribution along the cylinder.

The E-field is normalized to the maximum E-field near the focal point. 20 Hertzian dipoles are used for the validation. Due to the symmetry of the array, the elements from the center to the sides are driven with identical amplitude but different phase excitations. For the $\it{Ex}$ focusing scenario, the phase distribution of the elements is as follows: 0.56\textdegree, 5.06\textdegree, 14.05\textdegree, 27.51\textdegree, 45.42\textdegree, 67.74\textdegree, 94.44\textdegree, 125.47\textdegree, 160.77\textdegree, and 200.28\textdegree.

In the $\it{Ez}$ focusing scenario, the phase distribution on one side was modified to: 180.56\textdegree, 185.06\textdegree, 194.05\textdegree, 207.51\textdegree, 225.42\textdegree, 247.74\textdegree, 274.44\textdegree, 305.47\textdegree, 340.77\textdegree, and 20.28\textdegree. The simulation results were also normalized, as shown in Fig. 16 and Fig. 17, which compare the focusing performance of the electromagnetic simulation results.

The theoretical focal point is located at 20$\lambda$. The comparison results demonstrate that, aside from some minor differences in the intensity and position of the side lobes, the main lobes show a high degree of similarity. The discrepancies mainly arise from the differences between the simulation model and the ideal Hertzian dipole. The finite length of the dipole introduces a phase distribution along the dipole that is not perfectly uniform, which in turn affects the side lobes of the focused field.
 \begin{figure}
	\centerline{\includegraphics[width=3.6in]{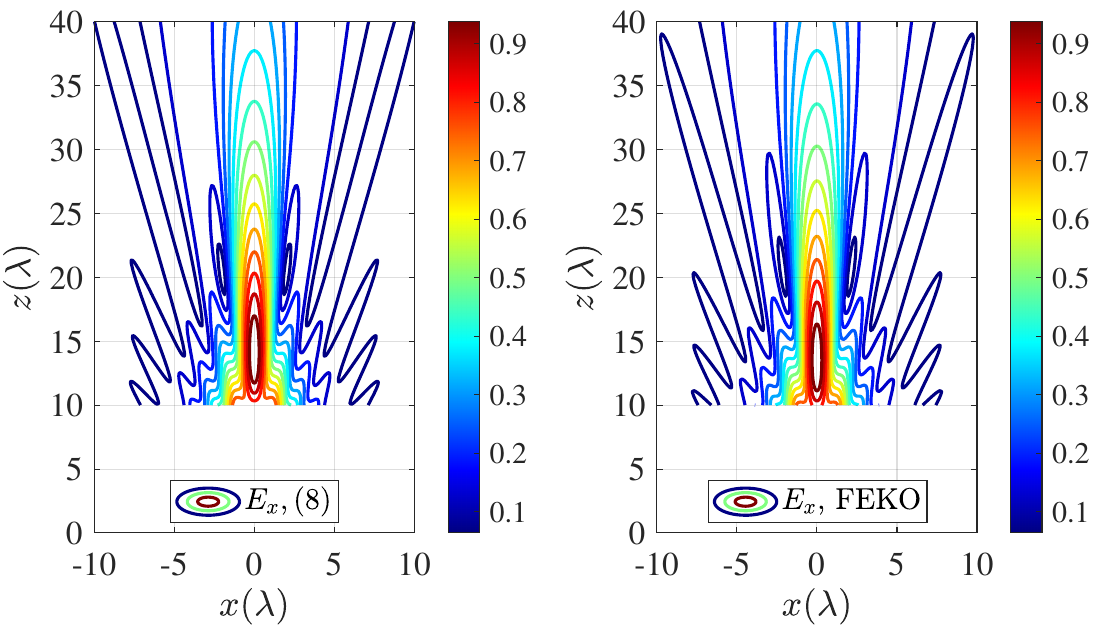}}
	\caption{Comparison of model simulation results with electromagnetic simulation results for $E_x$. \label{fig1}}
 \end{figure}
 
 \begin{figure}
	\centerline{\includegraphics[width=3.6in]{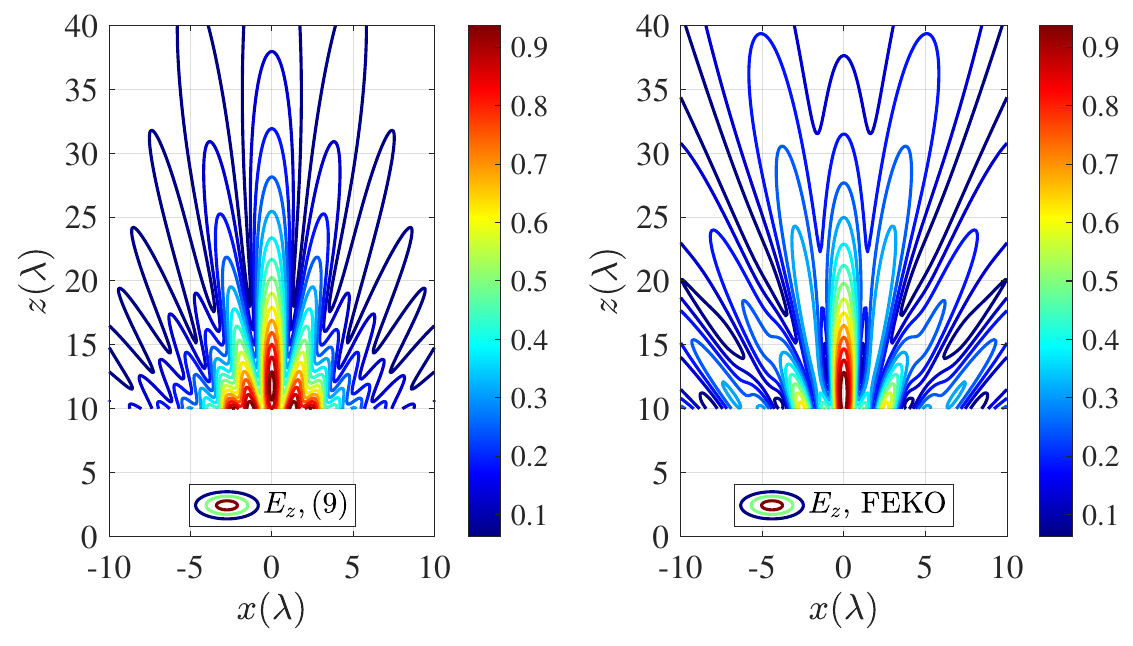}}
	\caption{Comparison of model simulation results with electromagnetic simulation results for $E_z$. \label{fig1}}
 \end{figure}
\section{CONCLUSION}
The study evaluates the 3dB focal resolution generated by the dipole array for both vertical and horizontal polarizations. Additionally, the feasibility of generating CP focal points and the focal displacement under a limited array size are explored.
Previous near-field focal synthesis methods primarily considered only the LOS channel. To enhance practicality, this study adopts a two-ray channel model to analyze near-field focal synthesis results in the presence of a reflected path. Horizontal and vertical polarizations are discussed separately, with design guidelines provided for array placement and focal point positioning.
Finally, electromagnetic simulations are conducted for both vertical and horizontal polarizations within the linear array to validate the proposed design. These analysis highlight the capability of the horizontal dipole array configuration to be directly applied without significant coupling effects between elements. The proposed design guidelines lay a solid foundation for the application of LIS in 6G technology.

\section*{APPENDIX}
The derivation process for the closed-form solutions of the beamwidth and beam depth for \(E_x\) and \(E_z\) is shown in (32) to (35). For the calculation of the focal width, this involves the gain variation along the array direction, which introduces a small increment in the variable \(s\). Conversely, the calculation of the focal depth reflects the gain variation along the dimension perpendicular to the array direction, achieved by introducing a small increment in the focal variable \(z_0\).
\begin{equation}
\begin{aligned}
E_x(\delta_w) = &\int_{-L/2}^{L/2} 
\frac{z_0^2 e^{-jk \left( \sqrt{z_0^2 + s^2} - \sqrt{z_0^2 + (s - \delta)^2} \right)}}
{\left(z_0^2 + (s - \delta)^2\right)^{3/2}} \, ds \\
\approx &\left| \int_{-L/2}^{L/2} 
\frac{z_0^2 e^{\frac{jk \delta}{\sqrt{z_0^2 + s^2}}}}
{\left(z_0^2 + s^2\right)^{3/2}} \, ds \right| \\
= &\left| 2 \sin \left(\frac{k \delta}{\sqrt{1 + 4 \left(\frac{z_0}{L}\right)^2}}\right) 
\big/ (k \delta) \right| \\
\approx &2 \left|\frac{\sin(k \delta)}{k \delta}\right| 
= 2 \left|\text{sinc}(k \delta)\right|,
\end{aligned}
\end{equation}

\begin{equation}
\begin{aligned}
E_x(\delta_d) = &\int_{-L/2}^{L/2} 
\frac{(z_0 + \delta_d)^2 e^{-jk \left( \sqrt{z_0^2 + s^2} - \sqrt{(z_0 + \delta_d)^2 + s^2} \right)}}
{\left((z_0 + \delta_d)^2 + s^2\right)^{3/2}} \, ds \\
\approx &\left| \int_{-L/2}^{L/2} 
\frac{z_0^2 e^{\frac{jk \delta_d}{\sqrt{z_0^2 + s^2}}}}
{\left(z_0^2 + s^2\right)^{3/2}} \, ds \right| \\
= &\left| 2 z_0^2 \int_{z_0}^{\infty} 
\frac{e^{-\frac{jk \delta_d}{t}}}
{t^2 \sqrt{t^2 - z_0^2}} \, dt \right| \\
= &\pi \left| j J_1(k \delta_d) + H_{-1}(k \delta_d) \right|.
\end{aligned}
\end{equation}
\begin{equation}
\begin{aligned}
E_z(\delta_w) = &\int_{-L/2}^{L/2} 
\frac{|s + \delta| z_0 e^{-jk \left( \sqrt{z_0^2 + s^2} - \sqrt{z_0^2 + (s + \delta)^2} \right)}}
{\left(z_0^2 + (s + \delta)^2\right)^{3/2}} \, ds \\
= &\int_{-L/2}^{L/2} 
\frac{|s + \delta| z_0 e^{\frac{jk \delta s}{\sqrt{z_0^2 + s^2}}}}
{\left(z_0^2 + (s + \delta)^2\right)^{3/2}} \, ds \\
\approx &z_0 \int_{-L/2}^{L/2} 
\frac{|s| e^{\frac{jk \delta s}{\sqrt{z_0^2 + s^2}}}}
{\left(z_0^2 + s^2\right)^{3/2}} \, ds \\
= &2\left| \int_{0}^{\pi/2} 
\sin \theta \cos (k \delta \sin \theta) \, d\theta \right| \\
= &\pi \left| H_{-1}(k \delta) \right|,
\end{aligned}
\end{equation}

\begin{equation}
\begin{aligned}
E_z(\delta_d) = &\int_{-L/2}^{L/2} 
\frac{|s| (z_0 + \delta_d) e^{-jk \left( \sqrt{z_0^2 + s^2} - \sqrt{(z_0 + \delta_d)^2 + s^2} \right)}}
{\left((z_0 + \delta_d)^2 + s^2\right)^{3/2}} \, ds \\
\approx &\left(z_0 + \delta_d\right) \left|\int_{-L/2}^{L/2} 
\frac{|s| e^{\frac{jk \delta_d s}{\sqrt{z_0^2 + s^2}}}}
{\left(z_0^2 + s^2\right)^{3/2}} \, ds \right| \\
= & (z_0 + \delta_d) \int_{z_0}^{\sqrt{(L/2)^2 + z_0^2}} 
\frac{e^{-\frac{jk \delta_d}{r}}}
{r^3} \sqrt{r^2 - z_0^2} \, dr \\
= & 2 (z_0 + \delta_d) \int_{z_0}^{\sqrt{(L/2)^2 + z_0^2}} 
\frac{e^{-\frac{jk \delta_d}{r}}}{r^2} \, dr \\
= &\frac{2 (z_0 + \delta_d)}{z_0} \int_{0}^{\pi/2} 
e^{jk \delta_d \cos \theta} \sin \theta \, d\theta \\
= &\frac{2 (z_0 + \delta_d)}{z_0} \left| i \frac{1-e^{i \delta k}}{\delta k} \right|.
\end{aligned}
\end{equation}

\section*{REFERENCES}

\def\refname{\vadjust{\vspace*{-1em}}} 


\begin{thebibliography}{00}
\bibitem{b1} S.  Hu,  F.  Rusek,  and  O.  Edfors,  “Beyond  massive  MIMO:  The 
potential of data transmission with Large Intelligent Surfaces,” \textit{IEEE 
Trans. Signal Process.}, vol. 66, no. 10, pp. 2746-2758, May 2018. 
\bibitem{b2} J. Li, Y. Zhu, Y. Hu and W. Hong, "A Miniaturized Dual-Polarized Active Phased Array Antenna for 5G Millimeter-Wave Applications," \textit{IEEE Antennas Wireless Propag. Lett.}, vol. 23, no. 2, pp. 618-622, Feb. 2024.
\bibitem{b3} W. Hong, K. -H. Baek and S. Ko, "A Millimeter-Wave 5G Antennas for Smartphones: Overview and Experimental Demonstration," \textit{IEEE Trans. Antennas Propag.}, vol. 65, no. 12, pp. 6250-6261, Dec. 2017.
\bibitem{b4} X. Li, G. Niu, B. Zhao and L. Yang, "Single-Cut Far-Field Pattern Determination of Polarized Linear Antenna Array via Quasi-Far-Field Mathematical Absorber Reflection Suppression," \textit{IEEE Trans. Antennas Propag.}, vol. 71, no. 3, pp. 2778-2783, March 2023.
\bibitem{b5} P. L. Garcia-Muller and A. G. Roederer., "A physical optics based plane wave spectrum approach to the analysis of finite planar antennas," \textit{IEEE Trans. Antennas Propag.}, vol. 40, no. 8, pp. 906-911, Aug. 1992.
\bibitem{b6} N. Mézières, L. Le Coq and B. Fuchs., "Phaseless Spherical Near-Field Antenna Measurements With Reduced Samplings," \textit{IEEE Trans. Antennas Propag.}, vol. 71, no. 9, pp. 7447-7456, Sept. 2023.
\bibitem{b7} F. Lisi, A. Michel and P. Nepa., "Synthesis of Near-Field Arrays Based on Electromagnetic Inner Products," \textit{IEEE Trans. Antennas Propag.}, vol. 71, no. 6, pp. 4919-4931, June 2023.
\bibitem{b8} H. R. Behjoo, A. Pirhadi and R. Asvadi., "Optimal Sampling in Spherical Near-Field Antenna Measurements by Utilizing the Information Content of Spherical Wave Harmonics," \textit{IEEE Trans. Antennas Propag.}, vol. 70, no. 5, pp. 3762-3771, May 2022.
\bibitem{b9} Y. Tak, J. Park and S. Nam., "Mode-Based Estimation of 3 dB Bandwidth for Near-Field Communication Systems," \textit{IEEE Trans. Antennas Propag.}, vol. 59, no. 8, pp. 3131-3135, Aug. 2011.
\bibitem{b10} J. Xu, L. You, G. C. Alexandropoulos, X. Yi, W. Wang and X. Gao., "Near-Field Wideband Extremely Large-Scale MIMO Transmissions With Holographic Metasurface-Based Antenna Arrays," \textit{IEEE Trans. Wireless Commun.}, vol. 23, no. 9, pp. 12054-12067, Sept. 2024.
\bibitem{b11} W. -S. Lee, K. -S. Oh and J. -W. Yu., "Distance-Insensitive Wireless Power Transfer and Near-Field Communication Using a Current-Controlled Loop With a Loaded Capacitance," \textit{IEEE Trans. Antennas Propag.}, vol. 62, no. 2, pp. 936-940, Feb. 2014.
\bibitem{b12} S. Lv, Y. Liu, X. Xu, A. Nallanathan and A. L. Swindlehurst., "RIS-Aided Near-Field MIMO Communications: Codebook and Beam Training Design," \textit{IEEE Trans. Antennas Propag.}, vol. 23, no. 9, pp. 12531-12546, Sept. 2024.
\bibitem{b13} R. González Ayestarán, G. León, M. R. Pino and P. Nepa., "Wireless Power Transfer Through Simultaneous Near-Field Focusing and Far-Field Synthesis," \textit{IEEE Trans. Antennas Propag.}, vol. 67, no. 8, pp. 5623-5633, Aug. 2019.
\bibitem{b14}  R. Hansen and L. Bailin, “A new method of near field analysis,” 
\textit{IEEE Trans. Antennas Propag.}, vol. 7, no. 5, pp. 458-467, Dec. 1959. 
\bibitem{b15} A.  Buffi,  P.  Nepa,  and  G.  Manara,  “Design  criteria  for  near-field-focused planar arrays,” \textit{IEEE Antennas Propag. Mag.}, vol. 54, no. 1, 
pp. 40-50, Feb. 2012. 
\bibitem{b16} P. Nepa and A. Buffi, “Near-field-focused microwave antennas: Near-
field shaping and implementation,” \textit{IEEE Antennas Propag. Mag.},  vol. 
59, no. 3, pp. 42-53, June 2017. 

\bibitem{b17}  R. Liu and K. Wu, “Antenna array for amplitude and phase specified 
near-field multifocus,” \textit{IEEE Trans. Antennas Propag.}, vol. 67, no. 5, 
pp. 3140-3150, May 2019.
\bibitem{b18} Y. F. Wu, Y. J. Cheng, S. S. Yao and Y. Fan, "Millimeter-Wave Near-Field-Focused Full 2-D Frequency Scanning Antenna Array With Height-Modulated-Ridge Waveguide," \textit{IEEE Trans. Antennas Propag.}, vol. 69, no. 5, pp. 2595-2604, May 2021.
\bibitem{b19} Y. F. Wu and Y. J. Cheng, "Proactive Conformal Antenna Array for Near-Field Beam Focusing and Steering Based on Curved Substrate Integrated Waveguide," \textit{IEEE Trans. Antennas Propag.}, vol. 67, no. 4, pp. 2354-2363, April 2019.
\bibitem{b20}J. Z. X. Huang and Y. J. Cheng, "Near-Field Pattern Synthesis for Sparse Focusing Antenna Arrays Based on Bayesian Compressive Sensing and Convex Optimization," \textit{IEEE Trans. Antennas Propag.}, vol. 66, no. 10, pp. 5249-5257, Oct. 2018.
\bibitem{b21}W. Chen, Z. Niu, Y. Ruan, Y. Ding and C. Gu, ”, "Structured Sparsity Learning with Group $\textit{L}_{1/2}$ Regularization for Efficient Synthesis of Near-Field Multi-Focus Sparse Planar Arrays," \textit{IEEE Antennas Wirel. Propag. Lett.}, vol. 66, no. 10, pp. 5249-5257, Oct. 2018.
\bibitem{b22}J. Z. X. Huang and Y. J. Cheng, "Synthesis of Sparse Near-Field Focusing Antenna Arrays With Accurate Control of Focal Distance by Reweighted $\textit{l}_1$ Norm Optimization," \textit{IEEE Trans. Antennas Propag.}, vol. 69, no. 5, pp. 3010-3014, May 2021.
\bibitem{b23}J. Li, Y. Hu, L. Xiang, W. Kong and W. Hong, "Broadband Circularly Polarized Magnetoelectric Dipole Antenna and Array for K-Band and Ka-Band Satellite Communications," \textit{IEEE Trans. Antennas Propag.}, vol. 70, no. 7, pp. 5907-5912, July 2022.
\bibitem{b24}R. Xu, J. . -Y. Li, J. Liu, S.. -G. Zhou, K. Wei and Z.. -J. Xing, "A Simple Design of Compact Dual-Wideband Square Slot Antenna With Dual-Sense Circularly Polarized Radiation for WLAN/Wi-Fi Communications," \textit{IEEE Trans. Antennas Propag.}, vol. 66, no. 9, pp. 4884-4889, Sept. 2018.
\bibitem{b25}J. L. Gómez-Tornero, D. Blanco, E. Rajo-Iglesias and N. Llombart. -J. Xing, "Holographic Surface Leaky-Wave Lenses With Circularly-Polarized Focused Near-Fields—Part I: Concept, Design and Analysis Theory," \textit{IEEE Trans. Antennas Propag.}, vol. 61, no. 7, pp. 3475-3485, July 2013.
\bibitem{b26}D. Blanco, J. L. Gómez-Tornero, E. Rajo-Iglesias and N. Llombart, "Holographic Surface Leaky-Wave Lenses With Circularly-Polarized Focused Near-Fields—Part II: Experiments and Description of Frequency Steering of Focal Length," \textit{IEEE Trans. Antennas Propag.}, vol. 61, no. 7, pp. 3486-3494, July 2013.
\bibitem{b27}Constantine A. Balanis, "Antenna Theory Analysis and Design, Fourth Edition," \textit{John Wiley \& Sons, Inc., Hoboken, New Jersey.}, Feb 2016.
\bibitem{b28}S. -J. Park, D. -H. Shin and S. -O. Park, "Low Side-Lobe Substrate-Integrated-Waveguide Antenna Array Using Broadband Unequal Feeding Network for Millimeter-Wave Handset Device," \textit{IEEE Trans. Antennas Propag.}, vol. 64, no. 3, pp. 923-932, March 2016.
\bibitem{b29}B. Y. Toh, R. Cahill and V. F. Fusco, "Understanding and measuring circular polarization," \textit{IEEE Trans. Education.}, vol. 46, no. 3, pp. 313-318, Aug. 2003.
\bibitem{b30}Andreas F. Molisch, "Wireless Communications," \textit{John Wiley \& Sons Inc} Dec. 2010.
\end{thebibliography}
\end{document}